\documentclass[11pt,letter]{article}

\usepackage{geometry}
\usepackage[utf8]{inputenc}
\usepackage[english]{babel}
\usepackage{lmodern}
\usepackage{microtype}

\usepackage{graphicx}
\usepackage{subfigure}
\usepackage{authblk}
\usepackage{hyperref}
\usepackage{xcolor}
\usepackage{standalone}
\usepackage{tikz,ifthen,calc}
\usepackage{enumitem}

\usepackage{tabularx}

\usepackage{amsmath,amsfonts,amssymb}
\usepackage{amsthm}
\usepackage{bbm}

\usepackage[capitalise]{cleveref}

\newcommand{\C}{\ensuremath{\mathbb{C}}}

\newcommand{\Z}{\ensuremath{\mathbb{Z}}}

\newcommand{\R}{\ensuremath{\mathbb{R}}}
\newcommand{\F}{\ensuremath{{\mathbb{F}}}}
\newcommand{\one}{\ensuremath{{\mathbbm{1}\hspace*{-0.5mm}}}}

\newcommand{\Ann}{\ensuremath{\text{\rm Ann}}}
\newcommand{\Per}{\ensuremath{\text{\rm Per}}}
\newcommand{\Ord}{\ensuremath{\text{\rm Ord}}}
\newcommand{\supp}{\ensuremath{\text{\rm Supp}}}

\newcommand{\A}{\ensuremath{\mathcal{A}}}

\newcommand{\Patt}[2]{\ensuremath{\mathcal{L}_{#2}(#1)}}
\newcommand{\Lang}[1]{\ensuremath{\mathcal{L}(#1)}}

\renewcommand{\vec}[1]{\mathbf{#1}}

\newtheorem{theorem}{Theorem}
\newtheorem*{theorem*}{Theorem}

\newtheorem{lemma}[theorem]{Lemma}
\newtheorem*{lemma*}{Lemma}

\crefname{coro}{Corollary}{Corollaries}
\newtheorem*{conjecture}{Conjecture}

\crefname{prop}{Property}{Properties}

\renewcommand{\vec}[1]{\mathbf{#1}}

\theoremstyle{definition}
\newtheorem{example}{Example}

\newcommand\restr[2]{{% we make the whole thing an ordinary symbol
  \left.\kern-\nulldelimiterspace % automatically resize the bar with \right
  #1 % the function
  \vphantom{\big|} % pretend it's a little taller at normal size
  \right|_{#2} % this is the delimiter
  }}

\title{Low-Complexity Tilings of the Plane}
\author[ ]{Jarkko Kari\thanks{Research supported by the Academy of Finland grant 296018}}
\affil[ ]{Department of Mathematics and Statistics, University of Turku, Finland}

\begin{document}

\maketitle

\thispagestyle{empty}

\begin{abstract}
\noindent
A two-dimensional configuration is a coloring of the infinite grid $\Z^2$ with finitely many colors. For a finite subset $D$ of $\Z^2$, the $D$-patterns of a configuration are the colored patterns of shape $D$ that appear in the configuration. The number of distinct $D$-patterns of a configuration is a natural measure of its complexity. A configuration is considered having low complexity with respect to shape $D$ if
the number of distinct $D$-patterns is at most $|D|$, the size of the shape. This extended abstract is a short review of an algebraic method
to study periodicity of such low complexity configurations.

\end{abstract}

%\keywords{Pattern complexity, periodicity, Nivat's conjecture, low complexity configurations, low complexity subshifts, commutative algebra, algebraic subshifts, domino problem}

%\begin{document}

%
%
\section{Introduction}

Commutative algebra provides powerful tools to analyze low complexity configurations, that is, colorings of the two-dimensional grid that have sufficiently low number of
different local patterns. If the colors are represented as numbers, the low complexity assumption implies that the configuration is a linear combination of its translated
copies. This condition can be expressed as an annihilation property under the multiplication of a power series representation of the configuration by a non-zero two-variate polynomial,
leading to the study of the ideal of all annihilating polynomials. It turns out that the ideal of annihilators is essentially a principal ideal
generated by a product of so-called line polynomials, i.e., univariate polynomials of two-variate monomials. This opens up the possibility to
obtain results on global structures of the configuration, such as its periodicity. We first proposed this approach in~\cite{icalp,icalpfull} to study Nivat's
conjecture. It led to a number of subsequent results~\cite{focs,karimoutot,cai,szabados}. In this presentation we review the main results without proofs --
the given references can be consulted for more details. We start by briefly recalling the notations and basic concepts.

\subsection{Configurations and periodicity}

A $d$-dimensional \emph{configuration}  over a finite alphabet $A$ is an assignment of symbols of $A$ on the infinite grid $\Z^d$. For any configuration $c\in A^{\Z^d}$ and any cell $\vec{u}\in\Z^d$, we denote by $c_{\vec{u}}$ the symbol that $c$ has in cell $\vec{u}$. For any vector $\vec{t}\in\Z^d$,  the \emph{translation} $\tau^{\vec{t}}$
by $\vec{t}$ shifts a configuration $c$ so that $\tau^{\vec{t}}(c)_{\vec{u}}=c_{\vec{u}-\vec{t}}$ for all $\vec{u}\in\Z^d$.
We say that $c$ is \emph{periodic} if $\tau^{\vec{t}}(c)=c$ for some non-zero $\vec{t}\in\Z^d$. In this case $\vec{t}$ is a \emph{vector of periodicity} and $c$ is also termed
\emph{$\vec{t}$-periodic}. We mostly consider the two-dimensional setting $d=2$. In this case, if there are
two linearly independent vectors of periodicity then $c$ is called \emph{two-periodic}. A two-periodic $c\in A^{\Z^2}$ has automatically
horizontal and vertical vectors of periodicity $(k,0)$ and $(0,k)$ for some $k> 0$, and consequently a vector of periodicity in every rational direction. A two-dimensional periodic configuration that is not two-periodic is called \emph{one-periodic}.

\subsection{Pattern complexity}

Let $D\subseteq\Z^d$ be a finite set of cells, a \emph{shape}. A \emph{$D$-pattern} is an assignment $p\in A^D$  of symbols in shape $D$. A
\emph{(finite) pattern} is a $D$-pattern for some finite $D$. Let us denote by $A^*$ the set of all finite patterns over alphabet $A$, where the dimension $d$ is
assumed to be known from the context. We say that a finite pattern $p$ of shape $D$ \emph{appears} in configuration $c$ if for some $\vec{t}\in\Z^d$
we have $\tau^{\vec{t}}(c)|_{D}=p$. We also say that $c$ \emph{contains} pattern $p$. For a fixed  $D$,
the set of $D$-patterns that appear in a configuration $c$ is denoted by $\Patt{c}{D}$. We denote by $\Lang{c}$ the set of all finite patterns that appear in $c$, i.e., the union of
$\Patt{c}{D}$ over all finite $D\subseteq\Z^d$.

The \emph{pattern complexity} of a configuration $c$ with respect to a shape $D$ is the number of $D$-patterns that $c$ contains. A sufficiently low pattern
complexity forces global regularities in a configuration. A relevant threshold happens when the pattern complexity is at most $|D|$, the number of cells
in shape $D$. Hence we say that $c$ has \emph{low complexity} with respect to shape $D$ if
$$|\Patt{c}{D}|\leq |D|.$$
We call $c$ a \emph{low complexity configuration} if it has low complexity with respect to some finite shape $D$.

\subsection{Nivat's conjecture}

The original motivation to this work is the famous conjecture presented by Maurice Nivat in his keynote address for the 25th anniversary of
the European Association for Theoretical Computer Science at ICALP 1997. It concerns two-dimensional configurations that have low complexity with respect to a rectangular shape.

\begin{conjecture}[\cite{nivat}]
Let $c\in A^{\Z^2}$ be a two-dimensional configuration. If $c$ has low complexity with respect to some rectangle $D=\{1,\ldots,n\}\times\{1,\ldots,m\}$ then $c$ is periodic.
\end{conjecture}
\noindent
The conjecture is still open but several partial and related results have been established. The best general bound was proved in~\cite{cyrkra} where it was
shown that for any rectangle $D$ the condition $|\Patt{c}{D}|\leq |D|/2$ is enough to guarantee that $c$ is periodic. This fact can also be proved using the algebraic approach~\cite{szabados}.

The analogous conjecture in dimensions higher than two fails, as does a similar claim in two dimensions
for many other shapes than rectangles~\cite{cassaigne}.
We return to Nivat's conjecture and our results on this problem in Section~\ref{sec:nivat}.

\subsection{Basic concepts of symbolic dynamics}

Let $p\in A^D$ be a finite pattern of shape $D$.
The set $[p]=\{c\in A^{\Z^d}\ |\ c|_{D}=p\}$ of configurations that have $p$ in domain $D$ is called
the \emph{cylinder} determined by $p$.
The collection of cylinders $[p]$ is a base of a compact topology on $A^{\Z^d}$, the \emph{prodiscrete} topology.
The topology is equivalently defined by a metric on $A^{\Z^d}$ where two configurations are close to each other if they agree
with each other on a large region around cell $\vec{0}$ -- the larger the region the closer they are. Cylinders are clopen in the topology: they are both open and closed.

A subset $X$ of $A^{\Z^2}$ is called a \emph{subshift} if it is closed in the topology
and closed under translations. By a compactness argument, every configuration $c$ that is not in $X$ contains a finite pattern $p$
that prevents it from being in $X$: no configuration that contains $p$ is in $X$. We can then as well  define subshifts  using forbidden patterns:
given a set $P\subseteq A^*$ of finite patterns we define
$$X_P=\{c\in A^{\Z^d}\ |\  \Lang{c}\cap P=\emptyset\},$$
the set of configurations that do not contain any of the patterns in $P$.
Set $X_P$ is a subshift, and every subshift is $X_P$ for some $P$.
If $X=X_P$ for some finite $P$ then $X$ is a \emph{subshift of finite type} (SFT).

In this work we are interested in subshifts that have low pattern complexity. For a subshift $X\subseteq A^{\Z^d}$ (or actually for any set $X$ of configurations) we define
its language $\Lang{X}\subseteq A^*$ to be the set of all finite patterns that appear in some element of $X$, that is, the union of
sets $\Lang{c}$ over all $c\in X$. For a fixed shape $D$, we analogously define
$\Patt{X}{D}=\Lang{X}\cap A^D$, the union of all $\Patt{c}{D}$ over $c\in X$. We say that $X$
has low complexity with respect to shape $D$ if $|\Patt{X}{D}|\leq |D|$.
For example, in Theorem~\ref{thm:focs3} we fix shape $D$ and a small set
$P\subseteq A^D$ of at most $|D|$ allowed patterns of shape $D$.
Then $X=X_{A^D\setminus P}=\{c\in A^{\Z^d}\ |\ \Patt{c}{D}\subseteq P \}$
is a low complexity SFT since $\Patt{X}{D}\subseteq P$ and $|P|\leq |D|$.

The \emph{orbit} of a configuration $c$ is the set ${\cal  O}(c) = \{\tau^{\vec{t}}(c)\ |\ \vec{t}\in\Z^2\ \}$ of all its translates, and
the \emph{orbit closure} $\overline{{\cal  O}(c)}$ of $c$ is the topological closure of its orbit. The orbit closure is a subshift,
and in fact it is the intersection of all subshifts that contain $c$. In terms of finite patters,
$c'\in \overline{{\cal  O}(c)}$ if and only if every finite pattern that appears in $c'$ appears also in $c$. Of course,
the orbit closure of  a low complexity configuration is a low complexity subshift.

A configuration $c$ is called \emph{uniformly recurrent} if for every $c'\in \overline{{\cal  O}(c)}$ we have
$\overline{{\cal  O}(c')} = \overline{{\cal  O}(c)}$. This is equivalent to  $\overline{{\cal  O}(c)}$ being
a \emph{minimal subshift} in the sense that it has no proper non-empty subshifts inside it. A classical result by Birkhoff on dynamical systems
implies that every non-empty subshift contains a minimal subshift, so there is a uniformly recurrent configuration
in every non-empty subshift~\cite{birkhoff}.

\subsection{Algebraic concepts}

To use commutative algebra we assume that $A\subseteq \Z$, i.e., the symbols in the configurations are integers. We also maintain the assumption that $A$ is finite.
We express
a $d$-dimensional configuration $c\in A^{\Z^d}$ as a formal power series  over $d$ variables $x_1,\dots x_d$ where the monomials address
cells in a natural manner $x_1^{u_1}\cdots x_d^{u_d} \longleftrightarrow (u_1,\dots ,u_d)\in\Z^d$, and the coefficients of the monomials in the power series
are the symbols at the corresponding cells.
Using the convenient vector notation $\vec{x}=(x_1,\dots x_d)$ we write
$\vec{x}^{\vec{u}}=x_1^{u_1}\cdots x_d^{u_d}$ for the monomial that represents cell
$\vec{u}=(u_1,\dots u_d)\in\Z^d$. Note that all our power series and polynomials are \emph{Laurent} as we allow negative as well as positive powers of variables.
Now the configuration $c\in\A^{\Z^d}$ can be coded as the formal power series
$$
c(\vec{x}) = \sum_{\vec{u}\in\Z^d} c_{\vec{u}}\vec{x}^{\vec{u}}.
$$
Because $A\subseteq \Z$ is finite, the power series $c(\vec{x})$ is \emph{integral} (the coefficients are integers) and \emph{finitary}
(there are only finitely many different coefficients). Henceforth we treat configurations as integral, finitary power series.

Note that the power series are indeed formal: the role of the variables is only to provide the position information on the grid. We may sum up two power series, or multiply a power series with a polynomial, but we never plug in any values in the variables.
Multiplying a power series $c(\vec{x})$ by a monomial $\vec{x}^{\vec{t}}$ simply adds $\vec{t}$ to the exponents of all monomials, thus producing the
power series of the translated configuration $\tau^{\vec{t}}(c)$. Hence the configuration $c(\vec{x})$
is $\vec{t}$-periodic if and only if $\vec{x}^{\vec{t}}c(\vec{x})=c(\vec{x})$, that is,
if and only if $(\vec{x}^{\vec{t}}-1)c(\vec{x})=0$, the zero power series. Thus we can express the periodicity of a configuration in terms of its \emph{annihilation} under
the multiplication with a \emph{difference binomial} $\vec{x}^{\vec{t}}-1$. Very naturally then we introduce
the \emph{annihilator ideal}
$$\Ann(c) = \{ f(\vec{x})\in \C[\vec{x}^{\pm 1}] ~|~ f(\vec{x})c(\vec{x})=0 \}$$
containing all the polynomials that annihilate $c$. Here we use the notation $\C[\vec{x}^{\pm 1}]$ for the set of Laurent polynomials with complex coefficients.
%Analogously we
%denote by $\C[[\vec{x}^{\pm 1}]]$ the set of Laurent power series with complex coefficients.
Note that  $\Ann(c)$ is indeed an ideal of the Laurent polynomial ring $\C[\vec{x}^{\pm 1}]$.

Our first observation relates the low complexity assumption to annihilators.
Namely, it is easy to see using elementary linear algebra that any low complexity configuration has at least some non-trivial annihilators:
\begin{lemma}[\cite{icalp}]
\label{th:low_complexity}
Let $c$ be a low complexity configuration.
Then $\Ann(c)$ contains a non-zero polynomial.
\end{lemma}

One of the main results of~\cite{icalp} states that if a
configuration $c$ is annihilated by a non-zero polynomial (e.g., due to low complexity) then it is automatically annihilated by a product of difference binomials.
\begin{theorem}[\cite{icalp}]
  \label{th:decompo}
  Let $c$ be a configuration annihilated by some non-zero polynomial.
  Then there exist pairwise linearly independent $\vec{t}_1, \ldots, \vec{t}_m\in\Z^d$ such that
  \[ (\vec{x}^{\vec{t}_1} - 1) \cdots (\vec{x}^{\vec{t}_m} - 1) \in \Ann(c) .\]
\end{theorem}
Note that if $m=1$ then the configuration is $\vec{t}_1$-periodic. Otherwise, for $m\geq 2$,
annihilation by $(\vec{x}^{\vec{t}_1} - 1) \cdots (\vec{x}^{\vec{t}_m} - 1)$ can be considered
a form of generalized periodicity.

%For a set $X\subseteq  A^{\Z^d}$ of configurations, we denote by $\Ann(X)$ the set of Laurent polynomials that annihilate all elements of $X$, and
%we call $\Ann(X)$ the annihilator ideal of $X$.

In the two-dimensional setting $d=2$ we find it sometimes  more convenient to work with the \emph{periodizer ideal}
$$\Per(c) = \{ f(\vec{x})\in \C[\vec{x}^{\pm 1}] ~|~ \mbox{ $f(\vec{x})c(\vec{x})$ is two-periodic } \}$$
that contains those two-variate Laurent polynomials whose product with configuration $c$ is two-periodic. Clearly also $\Per(c)$ is
an ideal of the Laurent polynomial ring $\C[\vec{x}^{\pm 1}]$, and we have $\Ann(c)\subseteq\Per(c)$. In the two-dimensional case we have a
very good understanding of the structure of the ideals $\Ann(c)$ and $\Per(c)$, see Theorems~\ref{thm:periodizer} and \ref{thm:annihilator} in
Section~\ref{sec:structure}.

\section{Contributions to Nivat's conjecture}
\label{sec:nivat}

In \cite{icalp} we reported an asymptotic result on Nivat's conjecture. The complete proof appeared in~\cite{icalpfull}.
Recall that the Nivat's conjecture claims -- taking the contrapositive of the original statement -- that every non-periodic configuration has
high complexity with respect to every rectangle. Our result states that this indeed holds for all sufficiently large rectangles:

\begin{theorem}[\cite{icalp,icalpfull}]
Let $c$ be a two-dimensional configuration that is not periodic. Then $\Patt{c}{D}>|D|$ holds for all but finitely many rectangles $D$.
\end{theorem}
Recall that Theorem~\ref{th:decompo} gives for a low complexity configuration
an annihilator of the form $(\vec{x}^{\vec{t}_1} - 1) \cdots (\vec{x}^{\vec{t}_m} - 1)$.
If $m=1$ then $c$ is periodic, so it is interesting to consider the cases of $m\geq 2$. Szabados proved in\cite{szabados}~that
Nivat's conjecture holds in the case $m=2$. Note that this case is equivalent to $c$ being the sum of two periodic configurations~\cite{icalp}.

\begin{theorem}[\cite{szabados}]
\label{thm:szabados}
Let $c$ be a two-dimensional configuration that has low complexity with respect to some rectangle. If $c$ is the sum of two periodic configurations
then $c$ itself is periodic.
\end{theorem}
We have also considered other types of configurations. Particularly interesting are uniformly recurrent configurations since they occur in all non-empty subshifts. Recently we proved
that they satisfy Nivat's conjecture, even when rectangles are generalized to other \emph{discrete convex shapes}. We call shape $D\subseteq \Z^2$ convex if $D=S\cap \Z^2$
for some convex set $S\subseteq \R^2$. In particular, every rectangle is convex.

\begin{theorem}[\cite{focs}]
\label{thm:focs}
Two-dimensional uniformly recurrent configuration that has low complexity with respect to a finite discrete convex shape $D$ is periodic.
\end{theorem}

The presence of uniformly recurrent configurations in subshifts then directly yields the following corollary.
%It then directly follows that any low complexity subshift necessarily contains periodic configurations.

\begin{theorem}[\cite{focs}]
\label{thm:focs2}
Let $X$ be a non-empty two-dimensional subshift that has low complexity with respect to a finite discrete convex shape $D$. Then $X$ contains a periodic configuration. In particular, the orbit closure of a configuration that has low complexity with respect to $D$ contains a periodic configuration.
\end{theorem}
Note that the periodic element in the orbit closure of $c$ means that $c$ contains arbitrarily large periodic regions.

The existence of periodic elements provides us with an algorithm to determine if a given low complexity SFT is empty. This is a classical argument by Hao Wang~\cite{wang}:
There is a semi-algorithm for non-emptyness of arbitrary SFTs, and there is a semi-algorithm for the existence of a periodic configuration in a two-dimensional SFT.
The latter semi-algorithm is based on the fact that if a two-dimensional SFT contains a periodic configuration then it also contains a two-periodic configuration, and these
can be effectively enumerated and tested. Now, since we know that a two-dimensional SFT that has low complexity with respect to a convex shape is either empty or contains
a periodic configuration, the two semi-algorithms together yield an algorithm to test emptyness.

\begin{theorem}[\cite{focs}]
\label{thm:focs3}
There is an algorithm that -- given a set of at most $|D|$ patterns $P\subseteq A^D$ over a two-dimensional convex shape $D$ -- determines whether
there exists a configuration $c\in A^{\Z^2}$ such that $\Patt{c}{D}\subseteq P$.
\end{theorem}

\section{Line polynomials and the structure of the annihilator ideal}
\label{sec:structure}

For a polynomial $f(\vec{x}) = \sum f_\vec{u} \vec{x}^{\vec{u}}$, we call $\supp(f) = \{ \vec{u}\in\Z^d ~|~ f_\vec{u}\neq 0 \}$ its
\emph{support}. A \emph{line polynomial}  is a polynomial with all its terms aligned on the same line: $f$ is a line polynomial
in direction $\vec{u}\in\Z^d$ if and only if $\text{supp}(f)$ contains at least two elements and $\text{supp}(f)\subseteq \Z\vec{u}$.
%$ \{n\vec{u}\ |\ n\in\Z\}$.
(Note that this definition differs slightly from the one in \cite{icalp,icalpfull} where the line containing the non-zero terms was not required to go
through the origin. The definitions are the same up to multiplication by a monomial, i.e. a translation.)
Multiplying a configuration by a line polynomial is a one-dimensional process: different discrete lines $\vec{v}+\Z\vec{u}$
in the direction $\vec{u}$ of the line polynomial get multiplied independently of each other.

Difference binomials $\vec{x}^{\vec{t}}-1$ are line polynomials so the special annihilator provided by Theorem~\ref{th:decompo} is a product of line polynomials. Annihilation by a difference binomial means periodicity -- and this fact generalizes to any line polynomial: a configuration that is annihilated by a line polynomial in direction $\vec{u}$
is $n\vec{u}$-periodic for some $n\in\Z$. This is due to the fact that the line polynomial annihilator specifies a linear recurrence along the discrete lines in direction $\vec{u}$.

The annihilator and the periodizer ideals of a configuration have particularly nice forms in the two-dimensional setting. Recall that
$\langle f\rangle=\{gf\ |\ g\in\C[\vec{x}^{\pm 1}]\}$ is the \emph{principal ideal}
generated by Laurent polynomial $f$. It turns out that a two-dimensional periodizer ideal is a principal ideal generated by a product of line polynomials.

\begin{theorem}[adapted from~\cite{icalpfull}]
\label{thm:periodizer}
Let $c$ be a two-dimensional configuration with a non-trivial annihilator. Then $\Per(c)=\langle f\rangle$ for a product $f=f_1\cdots f_m$ of some line polynomials $f_1,\dots ,f_m$.
\end{theorem}

By merging line polynomials in the same directions  we can choose
$f_i$ in the theorem above so that they are in pairwise linearly independent directions. In this case $m$, the number of line polynomial factors,
only depends on $c$. We denote $m=\Ord(c)$ and call it the \emph{order} of $c$. If $\Ord(c)=1$ then $c$ is periodic, and Theorem~\ref{thm:szabados} states that
the Nivat's conjecture is true among configurations of order two.

Theorem~\ref{thm:periodizer} directly implies a simple structure on the annihilator ideal: any annihilation of $c$ factors through the two-periodic configuration
$f_1\cdots f_m c$.

\begin{theorem}[\cite{icalpfull}]
\label{thm:annihilator}
Let $c$ be a two-dimensional configuration with a non-trivial annihilator. Then $\Ann(c)=f_1\cdots f_m H$ where $f_1,\dots ,f_m$ are line polynomials
and $H$ is the annihilator ideal of the two-periodic configuration $f_1\cdots f_m c$.
\end{theorem}

As pointed out above, if $c$ is annihilated by a line  polynomial then $c$ is periodic.
The structure of $\Per(c)$ and $\Ann(c)$ allows us to generalize this to other annihilators. If a two-dimensional configuration
$c$ is annihilated (or even periodized) by a polynomial
without any line polynomial factors then it follows from Theorem~\ref{thm:periodizer} that $\Per(c)$ is generated by polynomial $1$, that is, $c$ itself is
already two-periodic. Similarly, if $\Per(c)$ contains a polynomial whose line polynomial factors are all in a common direction then
$\Per(c)=\langle f\rangle$ is generated by a line polynomial $f$  in this direction, implying that $c$ has  a line polynomial annihilator and is therefore periodic.
Such situations have come up in the literature under the theme of covering codes on the grid~\cite{Axenovich}.

\begin{example}
Consider the problem of placing identical broadcasting antennas on the grid $\Z^2$ in such a way that each cell that does not contain an antenna receives broadcast from exactly $a$ antennas and every cell containing an antenna receives exactly $b$ broadcasts. Assume that $D\subseteq\Z^2$ is the shape of coverage by an antenna at the origin. Let us represent this broadcast range as the Laurent polynomial $f(\vec{x})=\sum_{\vec{u}\in D} \vec{x}^{\vec{u}}$. Let $c$ be a configuration
over $A=\{0,1\}$ where we interpret $c_{\vec{u}}=1$  as the presence of an antenna in cell $\vec{u}$. Now, $c$ is a solution to the antenna placement problem if and only if $f(\vec{x})c(\vec{x})$ is the power series $(b-a)c(\vec{x})+a\one(\vec{x})$ where $\one(\vec{x})$ is the constant one power series $\one(\vec{x})=\sum_{\vec{u}\in \Z^2} \vec{x}^{\vec{u}}$. Indeed, $(b-a)c(\vec{x})+a\one(\vec{x})$ has values $b$ and $a$ in cells containing and not containing an antenna, respectively. In other words, $c$ is
a valid placement of antennas if and only if multiplying $c(\vec{x})$ with polynomial
$f(\vec{x})-(b-a)$ results in the two-periodic configuration $a\one(\vec{x})$.
If $f(\vec{x})-(b-a)$ has no line polynomial factors then we know that this condition forces $c$ to be two-periodic. For example, if $D=\{(x,y)\ |\ |x|+|y|\leq 1\}$ so that each antenna only broadcasts to its own cell and the four neighboring cells, then $b-a\neq 1$ implies two-periodicity of any solution.
\qed
\end{example}

\section{Low complexity configurations in algebraic subshifts}

In~\cite{karimoutot} we considered low complexity configurations in algebraic subshifts where the alphabet $A$ is a finite field $\F_p$.
As Lemma~\ref{th:low_complexity} works as well in this setup, we have that every low complexity configuration $c$ is annihilated by a non-zero polynomial
$f\in \F_p[\vec{x}^{\pm 1}]$. We then have that $c$ is an element of the \emph{algeraic subshift} $S_f=\{ c\in A^{\Z^d}\ |\ fc=0\}$ of all configurations over $A=\F_p$
that are annihilated by $f$. So, to prove Nivat's conjecture it is enough to prove it for
elements of algebraic subshifts. Clearly $S_f$ is of finite type, defined by forbidden patterns of shape $D=-\supp(f)$.
We remark that the theory of this type of algebraically defined subshifts is well developed, see for example~\cite{schmidt}.

\begin{example}
\label{ex:ledrappier}
Let $A=\F_2$. The Ledrappier subshift (also known as the 3-dot system)  is $S_f$ for $f=1+x_1+x_2$. Elements of $S_f$ are the space-time diagrams
of the binary state XOR cellular automaton that adds to the state of each cell modulo 2 the state of its left neighbor.
\qed
\end{example}

While Lemma~\ref{th:low_complexity} works just fine over finite fields $\F_p$, Theorem~\ref{th:decompo} does not:
it is not true that every element of every algebraic subshift would be annihilated by a product of difference polynomials.
However, configurations over $\F_p$ can be also considered as configurations over $\Z$, without making calculations modulo $p$.
If a configuration $c$ over $\F_p$ has low complexity then it also has low complexity as a configuration over $\Z$, and thus in $\Z$ it has
a special annihilator
$(\vec{x}^{\vec{t}_1} - 1) \cdots (\vec{x}^{\vec{t}_m} - 1)$ provided by  Theorem~\ref{th:decompo}. Now, considering all calculations modulo $p$ we see that this special
annihilator is also an annihilator over $\F_p$. We conclude that even over $\F_p$, every low complexity configuration has an annihilator
that is a product of difference binomials.

\begin{example}
\label{ex:ledrappier2}
Let $c$ be a low complexity configuration in the Ledrappier subshift of Example~\ref{ex:ledrappier}. It is then annihilated by $f=1+x_1+x_2$ and by
some $g=(\vec{x}^{\vec{t}_1} - 1) \cdots (\vec{x}^{\vec{t}_m} - 1)$ that is a product of difference binomials.
Because $f$ does not have line polynomial factors while all irreducible factors of $g$ are line polynomials, we have that $f$ and $g$ do not have any
common factors. Replacing $x_2$ by $f-1-x_1$ in $g$, we can entirely eliminate variable $x_2$ from $g$,
obtaining a new annihilator $g'=g-f'f$ of $c$ having no occurrence of variable $x_2$. This annihilator $g'(x_1)$ is non-zero because $f$ and $g$ do not have common factors,
which implies that $c$ is horizontally periodic. We can repeat the same reasoning in the vertical direction, obtaining that $c$ is two periodic.
\qed
\end{example}

The reasoning in the example above can be generalized to other algebraic subshifts.

\begin{theorem}[\cite{karimoutot}]
Let $c$ be a low complexity configuration of an algebraic subshift $S_f$.
\begin{itemize}
\item If $f$ has no line polynomial factors then $c$ is two-periodic.
\item If all line polynomial factors of $f$ are in a common direction then $c$ is periodic.
\end{itemize}
\end{theorem}
Note that in the theorem there is no assumption about the low complexity shape $D$, so the applicability of the theorem is not restricted to rectangles or convex shapes.

\section{Conclusions and Perspectives}

There remains many open questions for future study. Obviously, the full version of Nivat's conjecture is still unsolved. Our Theorem~\ref{thm:focs} suggests that
perhaps periodicity is forced by the low complexity condition not only on rectangles but on other convex shapes as well, as conjectured by Julien Cassaigne
in~\cite{cassaigne}. In his examples of non-periodic low complexity configurations, the low complexity shape $D$ is always non-convex.
Moreover, all two-dimensional low complexity configurations that we know consist of periodic sublattices~\cite{cassaigne,karimoutot}.
For example, even lattice cells may form a configuration that is horizontally but not vertically
periodic while the odd cells may have a vertical but no horizontal period. The interleaved non-periodic configuration may
have low complexity with respect to a scatted shape $D$ that only sees cells of equal parity. We wonder if there exist any low complexity configurations
without a periodic sublattice structure.

Theorem~\ref{thm:szabados} proves Nivat's conjecture for configurations of order two. However, $\Ord(c)=2$ case is special in the sense that $c$ is then a sum of periodic configurations, that is,
finitary power series.
In general, any configuration with a non-trivial annihilator is a sum of periodic power series~\cite{icalp}, but already when  $\Ord(c)=3$ these power series
may be necessarily non-finitary~\cite{cai}. It seems then that proving Nivat's conjecture for configurations of order three would reflect the general case
better than the order two case.
We also remark that proving Nivat's conjecture (for all convex shapes) would render the results of Section~\ref{sec:nivat} obsolete.

There are also very interesting questions concerning general low complexity SFTs. By Theorem~\ref{thm:focs2}, a two-dimensional SFT that is low complexity with respect to a convex shape
contains periodic configurations. Might this be true for non-convex shapes as well~? If so, analogously to Theorem~\ref{thm:focs3}, this would
yield and algorithm to decide emptyness of general low complexity SFTs. What about higher dimensions~? We do not know of any aperiodic low complexity SFT in any dimension $d$ of the space.
The following example recalls a family of particularly interesting low complexity SFTs.

\begin{example}
A $d$-dimensional cluster tile is a finite subset $D\subseteq \Z^d$, and a co-tiler is a subset $C\subseteq \Z^d$ such that $C\oplus D=\Z^d$. Visually,
$C$ gives positions where copies of tiles $D$ can be placed so that every cell gets covered by exactly one tile.
Looking at the situation from an arbitrary covered cell $\vec{u}$, we see that $C$ is a co-tiler of $D$ if and only if the set
$\vec{u}-D$ contains precisely one element of $C$, for every $\vec{u}\in\Z^d$.
Representing a co-tiler $C$  as the indicator configuration $c_{\vec{u}}=1$ if $\vec{u}\in C$ and $c_{\vec{u}}=0$ if $\vec{u}\not\in C$,
we have that the set of valid co-tilers for tile $D$
is a low complexity SFT: The only allowed patterns of shape $-D$ are those that contain single $1$, and there are $|D|$ such patterns.

%Representing the tile $D$ as the
%Laurent polynomial $f(\vec{x})=\sum_{\vec{u}\in D} \vec{x}^{\vec{u}}$ and the co-tiler as the configuration $c(\vec{x})=\sum_{\vec{u}\in C} \vec{x}^{\vec{u}}$,
%both over alphabet $A=\{0,1\}$, the tiling condition becomes $f(\vec{x})c(\vec{x})=\one(\vec{x})$, the constant configuration of values 1.
%The set of valid co-tilers of tile $D$ is a low complexity SFT: The only allowed patterns of shape $D$ are those that contain single $1$, and there $|D|$ such patterns.

The periodic cluster tiling problem asks whether every tile that has a co-tiler also has a periodic co-tiler~\cite{LagariasWang,Szegedy}. This is a special case of
the more general question on arbitrary low complexity SFTs discussed above. The periodic cluster tiling problem was recently answered affirmatively in the
two-dimensional case~\cite{bhattacharya}. In~\cite{icalp} we gave a simple algebraic proof in any number of dimensions for the case -- originally handled in~\cite{Szegedy} --
where $|D|$ is a prime number.
\qed
\end{example}

Finally, the structure of the annihilator ideal is not known in dimension higher than two. We wonder how Theorem~\ref{thm:annihilator} might generalize to the
three-dimensional setting.

\bibliographystyle{abbrv}
\bibliography{biblio}

\begin{thebibliography}{10}

\bibitem{Axenovich}
M.~A. Axenovich.
\newblock On multiple coverings of the infinite rectangular grid with balls of
  constant radius.
\newblock {\em Discrete Mathematics}, 268(1):31 -- 48, 2003.

\bibitem{bhattacharya}
S.~{Bhattacharya}.
\newblock {Periodicity and decidability of tilings of $\mathbb{Z}^{2}$}.
\newblock preprint arXiv:1602.05738, Feb. 2016.

\bibitem{birkhoff}
G.~D. Birkhoff.
\newblock Quelques th\'eor\`emes sur le mouvement des syst\`emes dynamiques.
\newblock {\em Bulletin de la Soci\'et\'e Math\'ematique de France},
  40:305--323, 1912.

\bibitem{cassaigne}
J.~Cassaigne.
\newblock Subword complexity and periodicity in two or more dimensions.
\newblock In G.~Rozenberg and W.~Thomas, editors, {\em Developments in Language
  Theory. Foundations, Applications, and Perspectives. Aachen, Germany, 6-9
  July 1999}, pages 14--21. World Scientific, 1999.

\bibitem{cyrkra}
V.~Cyr and B.~Kra.
\newblock {Nonexpansive $\Z^2$-subdynamics and Nivat's Conjecture}.
\newblock {\em Transactions of the American Mathematical Society},
  367(9):6487--6537, Feb 2015.

\bibitem{focs}
J.~Kari and E.~Moutot.
\newblock {Decidability and Periodicity of Low Complexity Tilings}.
\newblock preprint arXiv:1904.01267, April 2019.

\bibitem{karimoutot}
J.~Kari and E.~Moutot.
\newblock Nivat's conjecture and pattern complexity in algebraic subshifts.
\newblock {\em Theoretical Computer Science}, 2019.

\bibitem{cai}
J.~Kari and M.~Szabados.
\newblock An algebraic geometric approach to multidimensional words.
\newblock In A.~Maletti, editor, {\em Algebraic Informatics}, pages 29--42,
  Cham, 2015. Springer International Publishing.

\bibitem{icalp}
J.~Kari and M.~Szabados.
\newblock {An Algebraic Geometric Approach to Nivat's Conjecture}.
\newblock In M.~M. Halld{\'{o}}rsson, K.~Iwama, N.~Kobayashi, and B.~Speckmann,
  editors, {\em Automata, Languages, and Programming - 42nd International
  Colloquium, {ICALP} 2015, Kyoto, Japan, July 6-10, 2015, Proceedings, Part
  {II}}, pages 273--285. Springer, 2015.

\bibitem{icalpfull}
J.~Kari and M.~Szabados.
\newblock {An Algebraic Geometric Approach to Nivat's Conjecture}.
\newblock preprint arXiv:1605.05929, May 2016.

\bibitem{LagariasWang}
J.~C. Lagarias and Y.~Wang.
\newblock Tiling the line with translates of one tile.
\newblock {\em Inventiones Mathematicae}, 124:341--365, 1996.

\bibitem{nivat}
M.~Nivat.
\newblock {Keynote address at the 25th anniversary of EATCS, during ICALP 1997,
  Bologna}, 1997.

\bibitem{schmidt}
K.~Schmidt.
\newblock {\em Dynamical systems of algebraic origin}.
\newblock Progress in mathematics. Birkh{\"a}user Verlag, 1995.

\bibitem{szabados}
M.~Szabados.
\newblock Nivat's conjecture holds for sums of two periodic configurations.
\newblock In A.~M. Tjoa, L.~Bellatreche, S.~Biffl, J.~van Leeuwen, and
  J.~Wiedermann, editors, {\em SOFSEM 2018: Theory and Practice of Computer
  Science}, pages 539--551. Springer International Publishing, 2018.

\bibitem{Szegedy}
M.~Szegedy.
\newblock Algorithms to tile the infinite grid with finite clusters.
\newblock In {\em 39th Annual Symposium on Foundations of Computer Science,
  FOCS '98, November 8-11, 1998, Palo Alto, California, USA}, pages 137--147.
  IEEE Computer Society, 1998.

\bibitem{wang}
H.~Wang.
\newblock {Proving theorems by pattern recognition -- II}.
\newblock {\em The Bell System Technical Journal}, 40(1):1--41, 1961.

\end{thebibliography}

\end{document}